\documentclass[preprint,12pt]{elsarticle}

\usepackage{amssymb}
\usepackage{hyperref}
\hypersetup{
	bookmarks=true,
	colorlinks=true,
	allcolors=blue
}
\usepackage{amsthm}
\usepackage{inputenc}
\usepackage{multirow}
\usepackage{overpic}
\usepackage{colortbl}
\usepackage{listings}
\usepackage{indentfirst}
\usepackage[version=4]{mhchem}
\usepackage{verbatim}
\usepackage{booktabs}
\usepackage{subcaption}

\usepackage{lineno}

\journal{Nuclear Inst. and Methods in Physics Research, A}

\begin{document}

\begin{frontmatter}

	\title{A Light Yield Enhancement Method Using Wavelength Shifter for the STCF EMC}

	\author{Zekun Jia}
	\author{Hanlin Yu}
	\author{Hongkun Mo}
	\author{Yong Song}
	\author{Zhongtao Shen}
	\author{Yunlong Zhang\corref{cor1}}
	\ead{ylzhang@ustc.edu.cn}
	\author{Jianbei Liu}
	\author{Haiping Peng}
	\cortext[cor1]{Corresponding author at: State Key Laboratory of Particle Detection and Electronics, University of Science and Technology of China,
		Hefei 230026, China}

	\address{State Key Laboratory of Particle Detection and Electronics, University of Science and Technology of
		China, Hefei 230026, China}

	\begin{abstract}
		Super Tau-Charm Facility
		(STCF) is a next-generation high luminosity electron-positron collider facility and is currently one of the major options for accelerator-based particle
		physics experiment in China. The crystal-based electromagnetic
		calorimeter (EMC) with undoped CsI is a major sub-system of the STCF spectrometer. To fulfill the increasing physics requirements on
		measurement precision and the suffering of radiation, improving the detected light yield is an important task of STCF EMC R\&D.
		This paper studies  a ``wavelength shifting in propagation" scheme for STCF EMC using the nanostructured organosilicon luminophore (NOL).
		The studies are performed by both Monte Carlo simulation and experimental tests. The light yield is proved
		to be improved by $159\:\%$in the experiment. Meanwhile, a study of the NOL's radiation hardness is carried out
		to verify the reliability of the NOL. No obvious degradation in the performance  is observed with the total ionization dose up to
		100\:krad, far beyond the total radiation dose of STCF with ten years of operation.

	\end{abstract}


\end{frontmatter}


\section{Introduction}%
\label{sec:introduction}

Super Tau-Charm Facility (STCF)~\cite{Luo:2019gri,snowmass} is a symmetric electron-positron collider covering the center-of-mass
energy ($\sqrt{s}$) from 2 to 7\:GeV and with a  peak luminosity larger than  $0.5\times10^{35}\:\mathrm{cm}^{-2}\mathrm{s}^{-1}$ optimized at  $\sqrt{s}=
	4\:\mathrm{GeV}$. STCF is
one of the major options for future accelerator-based particle physics experiments in China, and is in the technical R\&D stage.
The electromagnetic calorimeter (EMC) of STCF is required to detect photons with energy covering 0.025 to 3.5\:GeV
with high efficiency to conform to future physics studies. Specific to a 1\:GeV photon, STCF EMC is required to measure the energy with a precision of
$2.5\:\%$ and the position with a resolution of 6\:mm for its precision
kinematics requirement. It should also determine the signal time with a resolution of 300\:ps for the capability of particle identification among the neutral particles.

The luminosity of STCF will exceed the ultimate level of $0.5\times10^{35}\:\mathrm{cm}^{-2}\mathrm{s}^{-1}$. As a consequence, the beam-induced pile-up effect and radiation dose
are severe for STCF and challenge its manufacture and operation.
To cope with the requirements on precision and challenges from the high event rate and radiation dose, an undoped CsI-based EMC with avalanche photodiode
(APD) readout is proposed for the STCF spectrometer.
The undoped CsI and APD-based EMC features fast response and good radiation tolerance.
However, at the scene of undoped CsI directly coupled to APD, the effective detected light yield (L.Y.) is generally less than
50\:p.e.$/\mathrm{MeV}/\mathrm{cm}^{2}$~\cite{Prokhorova:2020jem,Jin:2016mic} (normalized by the effective area of APD)
. By utilizing the recently developed readout electronics for STCF EMC~\cite{luo2020design}, the expected equivalent noise energy is
$~$1.2\:MeV.
This feature significantly influences physics measurement precision. Therefore, increasing the effective
detected L.Y. is crucial and the priority for its design and manufacture.

In this paper, a ``wavelength shifting in propagation" (WLSP) scheme utilizing a novel nanostructured organosilicon
luminophore (NOL) is proposed to increase the effective detected L.Y. of undoped CsI-based EMC. The Monte Carlo (MC) simulation demonstrates the scheme and the cosmic
ray test validates it. Meanwhile,  studies of the radiation tolerance of the NOL are carried out.
The obtained results prove that the WLSP scheme can increase the effective detected L.Y. significantly for the undoped
CsI-based EMC by applying the NOL, and the radiation tolerance of the NOL is beyond more than ten years of operation of STCF.
The studies in this paper expand the use of wavelength
shifter in a sense and lay the foundation for manufacturing of the STCF EMC in the future.

\section{Concept of EMC module with wavelength shifter}%
\begin{figure}[htpb] \centering \includegraphics[width=\linewidth]{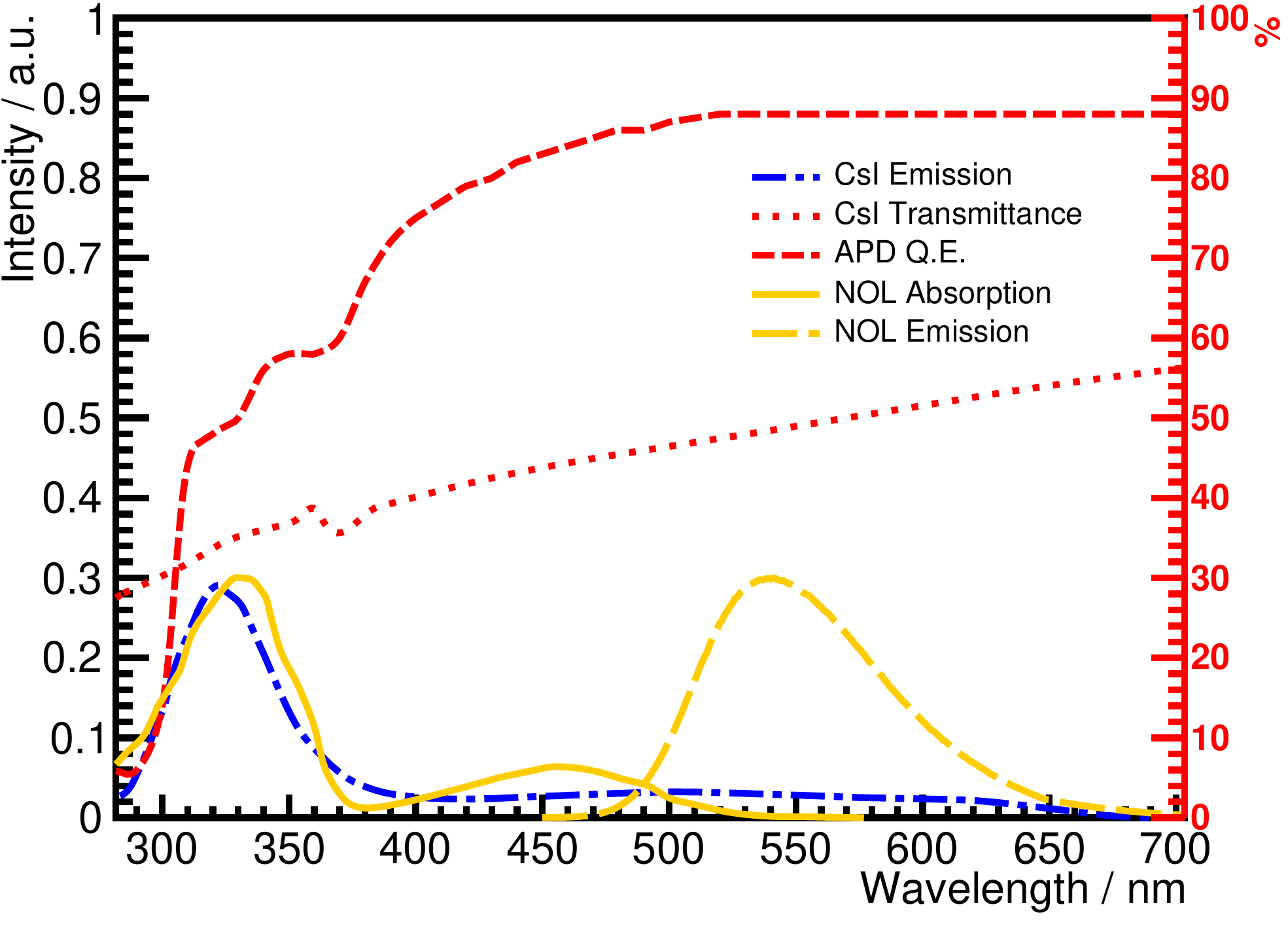}
	\caption{Properties of undoped CsI~\cite{ren}, APD~\cite{hamamatsu} and NOL. The blue dash-dotted line shows the emission spectrum of undoped CsI. The dotted red line shows the
		longitudinal transmittance of undoped CsI. The short-dashed red line shows the Q.E. of APD. The solid and long-dashed orange lines show the absorption and emission
		spectrum of NOL. The red axis on the right is for the dotted red and short-dashed red lines in the plot.}%
	\label{fig:nol9}
\end{figure}

\label{sec:basic_properties_of_undoped_csi_and_apd}
Undoped CsI is one of the fast scintillators with a decay time of 30\:ns and high radiation tolerance under a total-ionization dose (TID) up to several
dozens of\:krad~\cite{Bedny:2009zz,8456600}. It's considered the ideal absorber of the crystal-based EMC and widely used in particle physics
experiments, e.g.,  KTeV~\cite{Kessler:1995xz}, KOTO~\cite{Yamanaka:2012yma} and Mu2e~\cite{Atanov:2017mvq}
experiments. Therefore, STCF EMC will adopt the undoped CsI crystal as absorber, together with HAMAMATSU S8664 APD (denoted
as APD from now on) as its readout device due to its high quantum efficiency (Q.E.) and magnetic field-resistance compared to other photon detectors. The basic properties of undoped CsI,
i.e., emission and longitudinal transmittance spectrum, and the
Q.E. spectrum of APD are shown in Fig.~\ref{fig:nol9}.

However, the light yield (L.Y.) of undoped CsI is about 50 times smaller than that of \ce{CsI(Tl)}. Meanwhile, as shown in
Fig.~\ref{fig:nol9}, the major luminescence component of undoped CsI peaking
around 318\:nm (referred to as the UV component) has a low transmittance and APD Q.E., which
further degrades its intensity. Another component in the
visible wavelength range from 380\:nm to 700\:nm (from now on referred to as the VIS component) has a higher transmittance and Q.E., but the ratio of
VIS component intensity is only $\sim30\:\%$. The low crystal transmittance and APD Q.E. for the UV component heavily
limit the EMC energy and time measurement performance. So NOL, a wavelength shifter, is applied to mitigate the influence of both transmittance and
Q.E.

The NOL~\cite{10.1117/12.2187281,skorotetcky2019synthesis} to be used in the module is developed by LumInnoTech LLC, which has drawn
great interest in the field of high energy physics experiments~\cite{Jin:2016mic, borshchev2017development, akimov2017test, baxter2020coherent} in recent
years.
The NOL molecular branches two organic luminophore fragments together through silicon atoms. One of the fragments absorbs the photons with specific
wavelengths and excites the other fragment by transferring the generated electronic excitation energy via the silicon atoms. Then the excited fragment
de-excites and emits photons with a larger wavelength.
This approach can achieve a high quantum conversion rate and a specific absorption and emission spectra by finely tuning the characteristics of two fragments and the corresponding energy gap.
Figure~\ref{fig:nol9} shows the absorption and emission spectra for a specific NOL developed by LumInnoTech LLC, which has a peak absorption coefficient of
$7.4\times10^4\mathrm{cm}^2/\mathrm{g}$, a quantum yield of $95\:\%$ and a luminescence decay time of 7.17\:ns~\cite{luminotech}. The absorption spectrum of the
NOL matches well with the emission spectrum of undoped CsI. The emission spectrum of the NOL is of the wavelength above 500\:nm, which corresponds to high Q.E. for APD and high transmittance for undoped CsI.

\begin{figure}[htpb]
	\centering
	\includegraphics[width=0.9\linewidth]{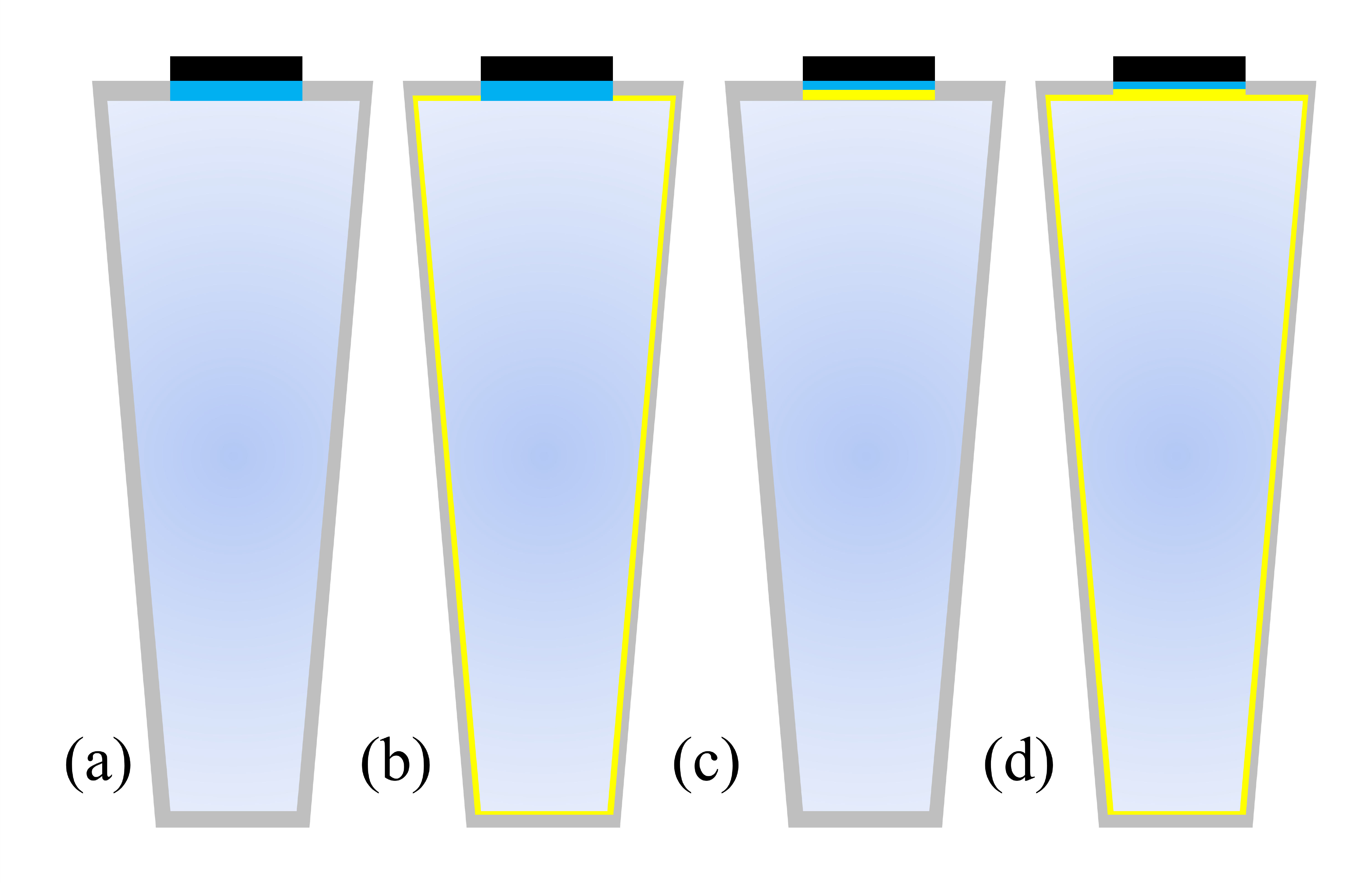}
	\caption{Layout of EMC modules with different NOL coating: (a) no-coating scheme; (b) WLSP scheme; (c) alternative-1 scheme; (d) alternative-2 scheme.
		The wedge-shaped crystal (light blue colored) is fully wrapped by a crystal package (gray colored).
		APD (black colored) is attached to the back end of the module, mediated by silicone grease (blue colored). The yellow areas highlight the NOL
		coating.}%
	\label{fig:schemes}
\end{figure}

The layout of the ordinary scheme (denoted as the no-coating scheme from now on) is shown in  Fig.~\ref{fig:schemes}(a). In general, the crystal module is
wedge-shaped, with the front and back ends as
a square of sizes ${5\times 5}\:\mathrm{cm}^2$ and  ${6\times 6}\:\mathrm{cm}^2$, respectively, and a length of 28\:cm corresponding to fifteen times of radiation length.
A $2\times2$ APD array with a total size of $2\times 2\:\mathrm{cm}^2$ is attached on the back end of the module, mediated by silicone grease with a thickness of
$10\:\mathrm{\mu m}$. Two layers of Tyvek fully wrap the crystal with a total thickness of $200\:\mathrm{\mu m}$, except for the area covered
by APDs. It's worth noting that this work uses Tyvek as the crystal package for the convenience of reassembly.
The electronics' signals of APDs are read out based on the charge sensitive amplifier (CSA)~\cite{luo2020design}.

Schemes utilizing NOL are shown in Fig.~\ref{fig:schemes} (b)(c)(d), which are named as ``wavelength shifting in propagation" (WLSP) scheme, alternative-1
scheme, and
alternative-2 scheme, respectively. The WLSP scheme coats a $50\:\mathrm{\mu m}$ NOL film on the Tyvek, which could reach the peaking absorption efficiency closing
to $100\:\%$~\cite{luminotech}. After wrapping the Tyvek on the crystal, the UV component photons are
converted into wavelengths above 500\:nm with high efficiency when reaching the surface between the crystal and Tyvek. This method is expected to improve the effective transmittance and
Q.E. of UV component simultaneously and the detected L.Y. eventually. This work mainly focuses on the WLSP scheme and will comprehensively study it in simulation and experiment.

For comparison, alternative-1 scheme
introduces NOL film between the APD and crystal, and alternative-2 scheme combines the above. Alternative-1 scheme can only improve the
effective Q.E. of the UV component and thus is limited by the low transmittance of crystal while alternative-2 scheme is expected to improve very little
when compared to the
WLSP scheme. Comparison between four schemes will be introduced in Section~\ref{sec:study_of_the_light_propagation_in_undoped_csi}.

\section{Simulation study of WLSP scheme for undoped CsI}%
\label{sec:study_of_the_light_propagation_in_undoped_csi}
\subsection{Setup of optical and cosmic ray simulation}%
\label{sub:simulation_setup_and_cosmic_ray_test_results}

\begin{figure}[htpb]
	\centering
	\includegraphics[width=\linewidth]{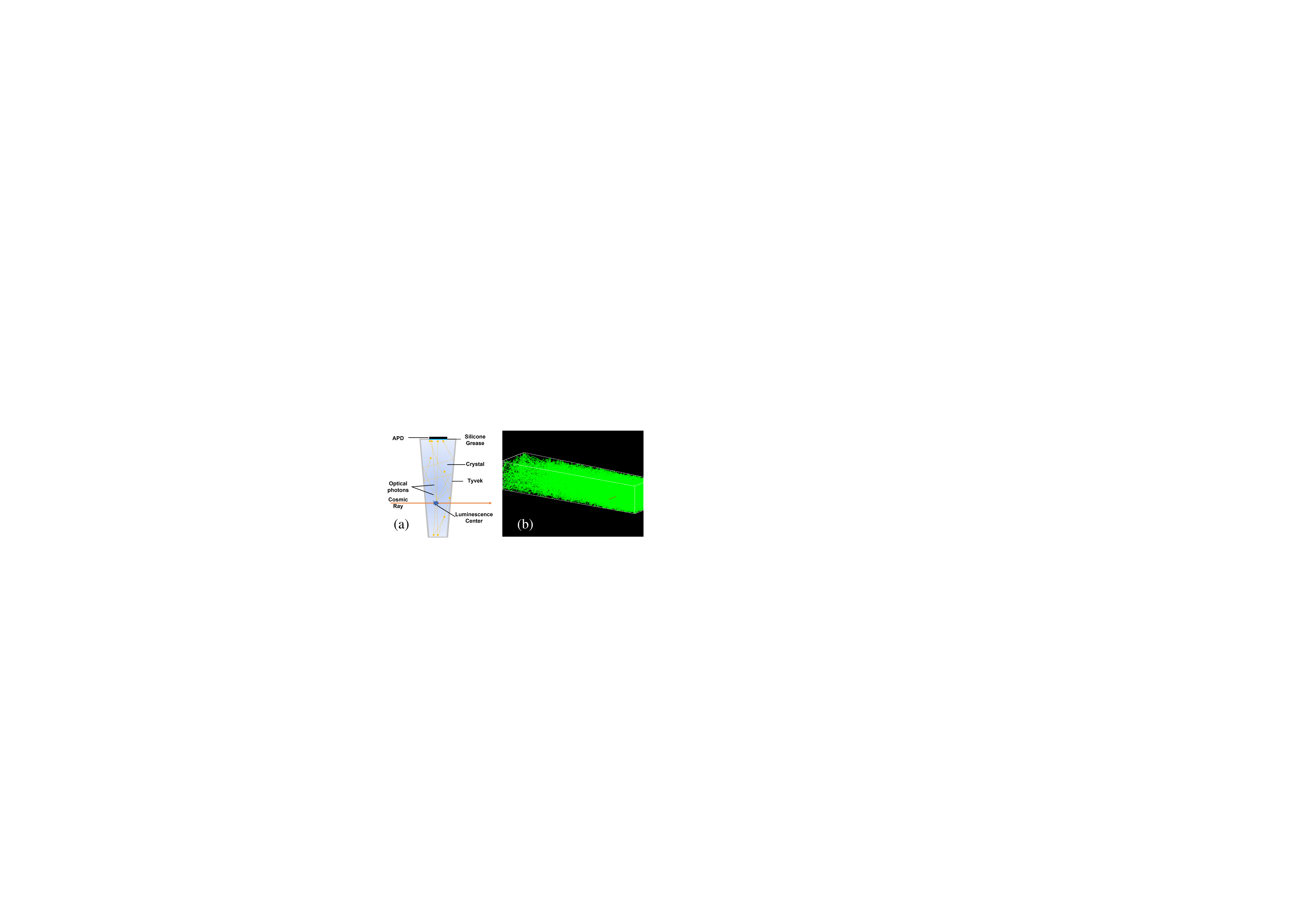}
	\caption{(a) Layout of the detector geometry in side view. The detector is composed of undoped CsI, Tyvek, silicone grease and APD. When a
		particle, e.g., the muon in the cosmic ray, interacts with the detector, the luminescence center is excited and generates isotropic scintillation photons. The photons are
		collected by APD after propagation in the crystal. (b) Illustration of a cosmic ray event. The gray lines outline the
		detector geometry. The red line is the incident cosmic ray. The green lines are the tracks of scintillation photons}%
	\label{fig:layout}
\end{figure}

The process of the light propagation in undoped CsI is simulated with the {\sc geant4}~\cite{GEANT4:2002zbu,Allison:2006ve,Allison:2016lfl} package.
In the simulation, besides the interaction between the incident particles and materials, the optical simulation is performed by incorporating various
parameters, i.e., the absorption length, emission spectrum~\cite{ren}, and refractive index~\cite{doi:10.1063/1.555536} of undoped CsI; the Q.E. of
APD~\cite{Ikagawa:2003md}; the diffuse reflectivity of Tyvek~\cite{6168236}, from the corresponding references. Besides, the  surface properties between
the crystal and Tyvek are parameterized by the UNIFIED model, and those between the APD and crystal are modeled by the Detector\_LUT model~\cite{Roncali_2017}
in {\sc geant4}. The wavelength shifting process is added to the simulation for schemes using NOL as a wavelength shifter. The parameters of NOL, including
the Q.E., emission and absorption spectra, absorption length and decay time, are obtained from Ref.~\cite{luminotech}.

The cosmic ray simulation is performed by shooting the cosmic ray muons  on the side face of the crystal using the CRY generator~\cite{4437209} to compare
with the experimental results of cosmic ray tests.
Figure~\ref{fig:layout}(a) shows the process of particle-matter interaction, scintillation generation, and signal detection.
Figure~\ref{fig:layout}(b) illustrates a specific simulated event.

\subsection{Performance of the WLSP scheme}%
\label{sec:performance_of_the_wlsp_scheme}

The simulation result is first checked for the no-coating scheme. The photons detected by APDs are counted individually within a time window of 200\:ns for the UV and VIS components.
The ratio of the UV and VIS components for generated and detected photons, as well as their average absorption lengths and Q.E. (weighted by the
wavelength spectrum), are summarized in  Table~\ref{tab:opsim}. From the table, the UV component accounts for $70\:\%$ in total and is dominant in generation. However,
its proportion degenerates to $32\:\%$ in total when detected by APD due to the smaller transmittance (or absorption length) and average Q.E.
\begin{table}[htpb]
	\centering
	\caption{Summary of the optical simulation parameters and result. }
	\label{tab:opsim}
	\begin{tabular}{ccc}
		\toprule
		Component              & UV    & VIS   \\
		\midrule
		Generated Ratio ($\%$) & 70    & 30    \\
		Absorption Length (cm) & 26.7  & 55.0  \\
		Q.E. ($\%$)            & 48    & 89    \\
		Detected Ratio ($\%$)  & 31.53 & 68.47 \\
		\bottomrule
	\end{tabular}
\end{table}

\begin{figure}[htpb]
	\centering
	\includegraphics[width=\linewidth]{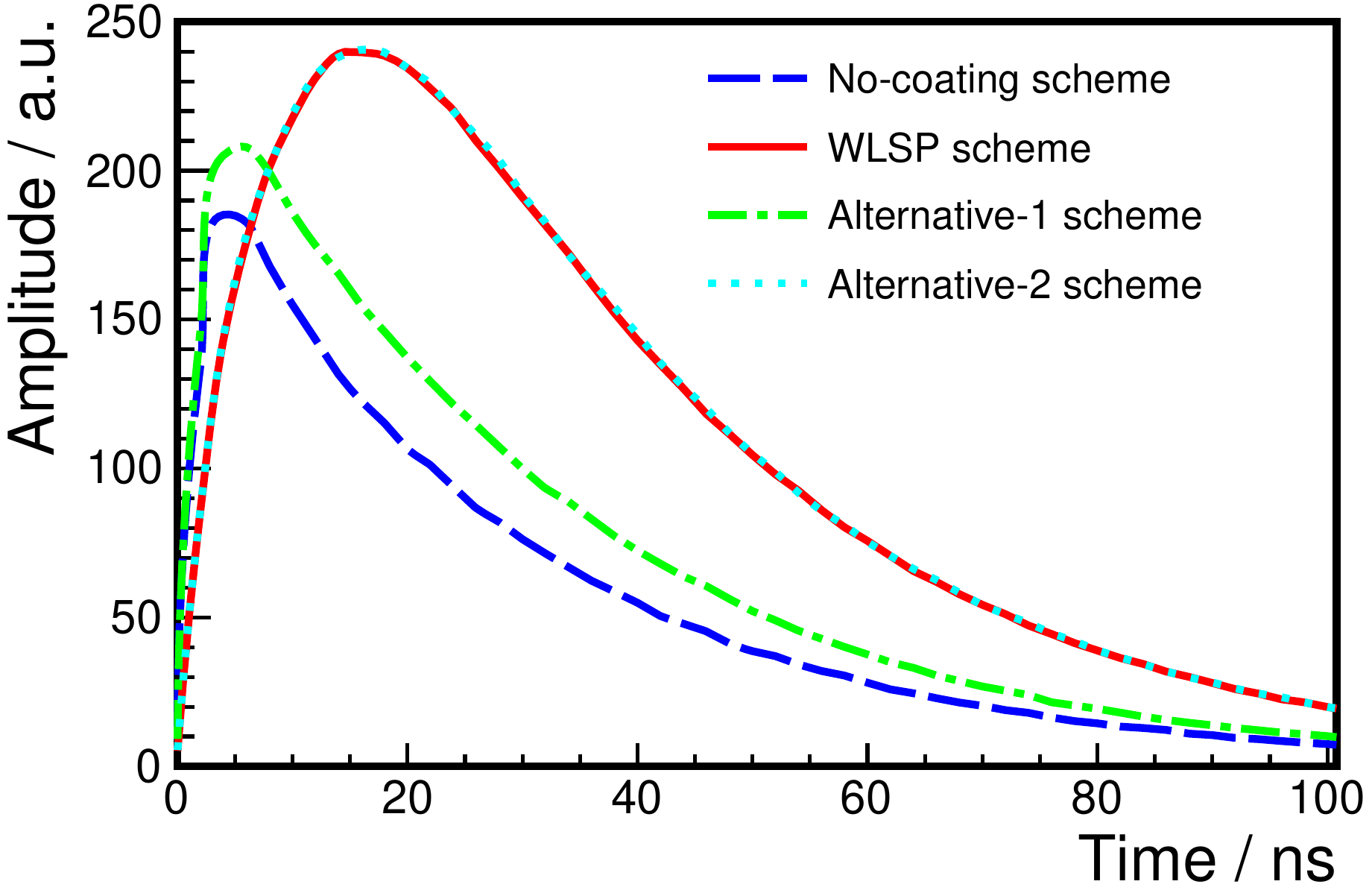}
	\caption{Waveform of the simulated cosmic ray signal. The dotted blue line is from
		no-coating scheme, the solid red line is obtained from WLSP scheme, the green dash-dotted line shows the result of alternative-1 scheme and the
		cyan dotted line is obtained from alternative-2 scheme.}%
	\label{fig:waveform}
\end{figure}

The other three schemes, especially the WLSP scheme, are then studied. The simulated waveforms read out by APDs in four schemes for a typical event are compared in Fig.~\ref{fig:waveform}. The area of the
waveform with the NOL is much larger than without the NOL, indicating an improvement in the detected L.Y. Still, the corresponding leading edge of the waveform
is a bit slower. This is induced by the 7.17\:ns decay time of NOL, which is kind of equivalent to convolving an exponential decay function on the
waveform of no-coating scheme. As the time is measured for the EMC of STCF with the output signal of the CSA~\cite{Luo:2022rjf}, the effects of the moderate leading edge on the time
resolution need further investigation.

Table~\ref{tab:sum_ly} summarizes the obtained L.Y. of four schemes in cosmic ray simulation. The simulation results show that the detected L.Y. is
improved by $136\:\%$ for the WLSP scheme, and by $25\:\%$ and $138\:\%$ for the alternative-1 and alternative-2 schemes, respectively.
This confirms the superiority of the WLSP method for undoped CsI: the effect of the WLSP scheme is much better than the alternative-1
scheme. This only improves the effective Q.E. of APD. On the other hand, the photon conversion efficiency is high enough in propagation; thus the alternative-2
scheme shows a similar L.Y. as the WLSP scheme.
\begin{table}[htpb]
	\centering
	\caption{Summary of the simulated light yield for four different schemes.}
	\label{tab:sum_ly}
	\begin{tabular}{ccc}
		\toprule
		Scheme               & L.Y. (p.e.$/\mathrm{MeV}$) & Relative Ratio \\
		\midrule
		No-coating scheme    & 143                        &                \\
		WLSP scheme          & 338                        & 2.36           \\
		Alternative-1 scheme & 179                        & 1.25           \\
		Alternative-2 scheme & 341                        & 2.38           \\

		\bottomrule
	\end{tabular}
\end{table}

\subsection{From cosmic ray test to collider experiment}%
\label{sub:from_cosmic_ray_test_to_collider_experiment}
In practice for collider experiments,  photons or electrons inject into the EMC from the front end of the crystal. Therefore, this section introduces further MC simulation
of the photon events to investigate the corresponding effects of the WLSP scheme.

The simulation setup is the same as that in Sec.~\ref{sub:simulation_setup_and_cosmic_ray_test_results}, only replacing the muon injecting from the side
face of the crystal with a 200\:MeV photon
injecting from the front end of the crystal  perpendicularly. The simulation results show that the average L.Y.s for the cases of the no
coating and the WLSP scheme are 180.2\:p.e./MeV and 341.5\:p.e./MeV for the injected
photons, respectively. The significant improvement of the detected L.Y. verifies the advantage of the WLSP scheme in practice of collider experiments. the average L.Y. for the photon incident (340 p.e./MeV) is very close to that in the cosmic ray test (338
p.e./MeV) in the case for the WLSP scheme. In this scenario, the effective transmittance becomes large enough and its dependence on the position
distribution of energy deposition gets small.

\section{Experiment validation of the WLSP scheme }%
\label{sec:cosmic_ray_test_of_the_wls_coating_emc_module}

In this section, an undoped CsI-based EMC module is manufactured to verify the effects of the WLSP scheme with the cosmic ray test.

\begin{figure}[htpb]
	\centering
	\includegraphics[width=\linewidth]{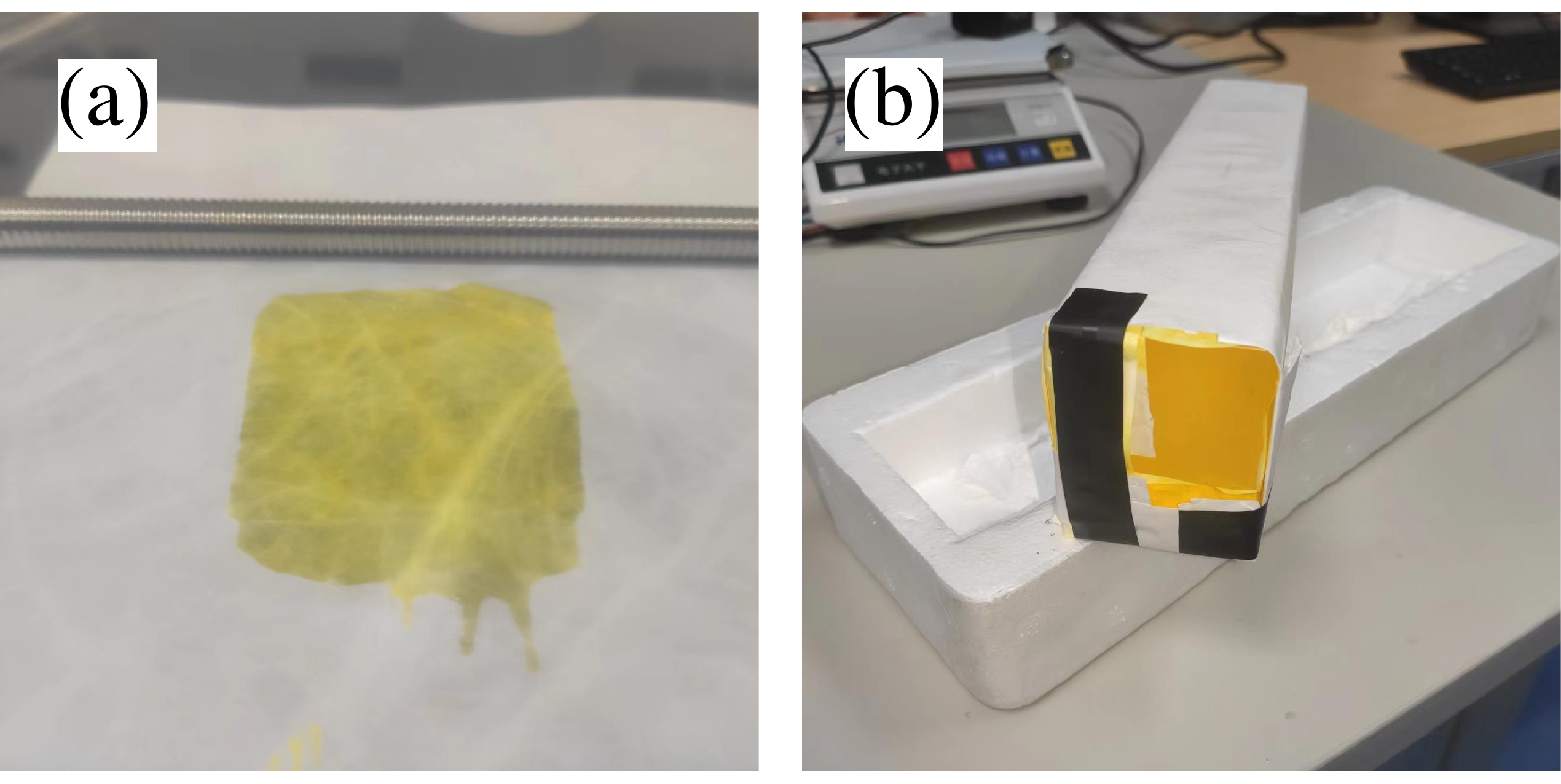}
	\caption{(a) a piece of Tyvek film coated with a Mayer bar and (b) the assembled module.}%
	\label{fig:pic_coat}
\end{figure}

The EMC module is manufactured with undoped CsI with the same size as described in Sec.~\ref{sub:simulation_setup_and_cosmic_ray_test_results}.
Four APDs with a size of  $1\times1\:\mathrm{cm}^2$ are attached on the back end of the crystal, and the scintillation signal is read out by APDs and the CSA-based electronics~\cite{luo2020design}.
The cosmic ray tests are carried out on a piece of crystal with two scenarios of crystal packing individually to verify the effects of the WLSP scheme, i.e., wrapping the
crystal with two layers of Tyvek with a total thickness of  $200\:\mathrm{\mu m}$ (corresponding to the no-coating scheme in simulation), and further
coating the NOL film on the inner layer of Tyvek as an NOL composite film (corresponding to the WLSP scheme in simulation).
The process of manufacturing an NOL composite film is as follows:
\begin{itemize}
	\item Dissolving the powder of the NOL in the toluene under the proportion of 0.4\:g/ml.
	\item Coating the solution on Tyvek film with the Mayer bar (as shown in  Fig.~\ref{fig:pic_coat} (a)). The wet thickness is $200\:\mathrm{\mu m}$.
	\item Placing the film in the fume hood until the film dries. After drying, the total thickness of the NOL composite film is $145\:\mathrm{\mu m}$.
\end{itemize}

\begin{figure}[htpb]
	\centering
	\includegraphics[width=\linewidth]{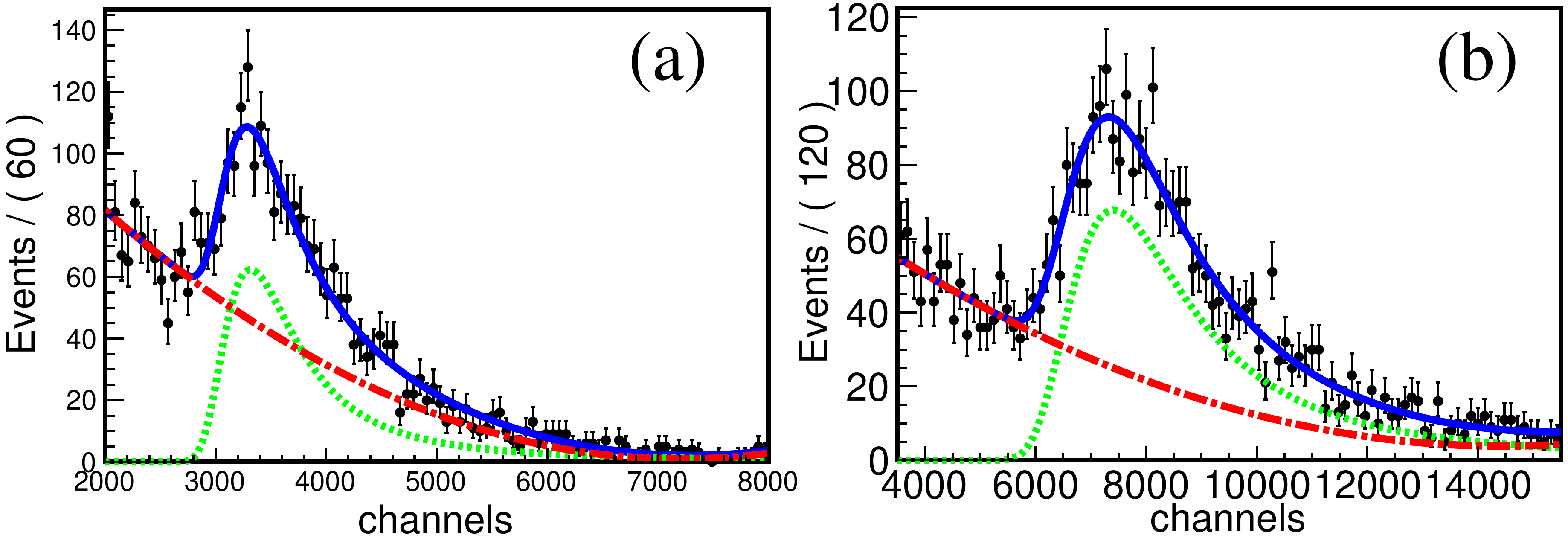}
	\caption{Energy spectrum acquired in the cosmic ray test of (a) no-coating and (b) WLSP scheme. The spectrum is fitted by a Landau distribution (dotted green line) and the second order polynomial
		(dash-dotted red line). The blue line shows the overall fit result.}%
	\label{fig:cr_spec}
\end{figure}

An EMC module wrapped with the NOL composite film is shown in Fig.~\ref{fig:pic_coat} (b). Triggered by the internal over-threshold discrimination of the
electronics, the data acquisition of the cosmic ray test is performed for around 24 hours each run.
Figure~\ref{fig:cr_spec} shows the energy spectrum acquired in the cosmic ray test for the scenarios without and with the NOL composite film individually, where the cosmic ray signal is observed clearly.
The energy spectrum is fitted with a Landau function representing the cosmic ray signal and the second order polynomial function
representing the non-peaking contribution. The average detected L.Y. of the cosmic ray is extracted with the peak position of the Landau function subtracting the
electronics pedestal.

The alternative-1 scheme is also validated experimentally by mounting a NOL-coated PMMA plate between the back facet of the crystal and APDs~\cite{Aihara:2016ct}.
As summarized in Table~\ref{tab:cr_result}, the obtained detected L.Y. of the scenario with the NOL composite film is improved by
$159\:\%$ relative to that of the scenario without the NOL composite film, and the L.Y. of alternative-1 scheme is only improved by
$\sim27\:\%$. This result is consistent with the simulation result shown in
Sec.~\ref{sec:performance_of_the_wlsp_scheme}, indicating the high wavelength shifting effiency of WLSP scheme in experiment.

\begin{table}[htpb]
	\centering
	\caption{Summary of the cosmic ray test results.}
	\label{tab:cr_result}
	\begin{tabular}{cccc}
		\toprule
		Scheme           & No-coating & WLSP   & Alternative-1 \\
		\midrule
		Pedestal channel & 724.5      & 729.7  & 722.0         \\
		Peak channel     & 3362.6     & 7561.4 & 4072.4        \\
		Signal amplitude & 2638.1     & 6831.7 & 3350.4        \\
		Amplitude ratio  & ---        & 2.59   & 1.27          \\
		\bottomrule
	\end{tabular}
\end{table}

\section{Radiation hardness test on the NOL}%
\label{sec:radiation_hardness_test_on_the_nol}

In the above sections, the NOL composite film coating the NOL has been verified to improve the detected L.Y. significantly. However, its radiation harness
is crucial for its application in the new generation experiments in practice and has not been explored yet.
In the case of STCF, the radiation of the EMC is dominated by the ionizing dose, which is expected to be up to a TID of $\sim$50\:krad in ten years.
So the radiation hardness test on the NOL composite film is carried out at the \ce{^{60}Co} $\gamma$-ray
radiation field (20\:kCi, located in the University of Science and Technology of China)~\cite{jiao2021effect}.

\begin{figure}[htpb]
	\centering
	\includegraphics[width=\linewidth]{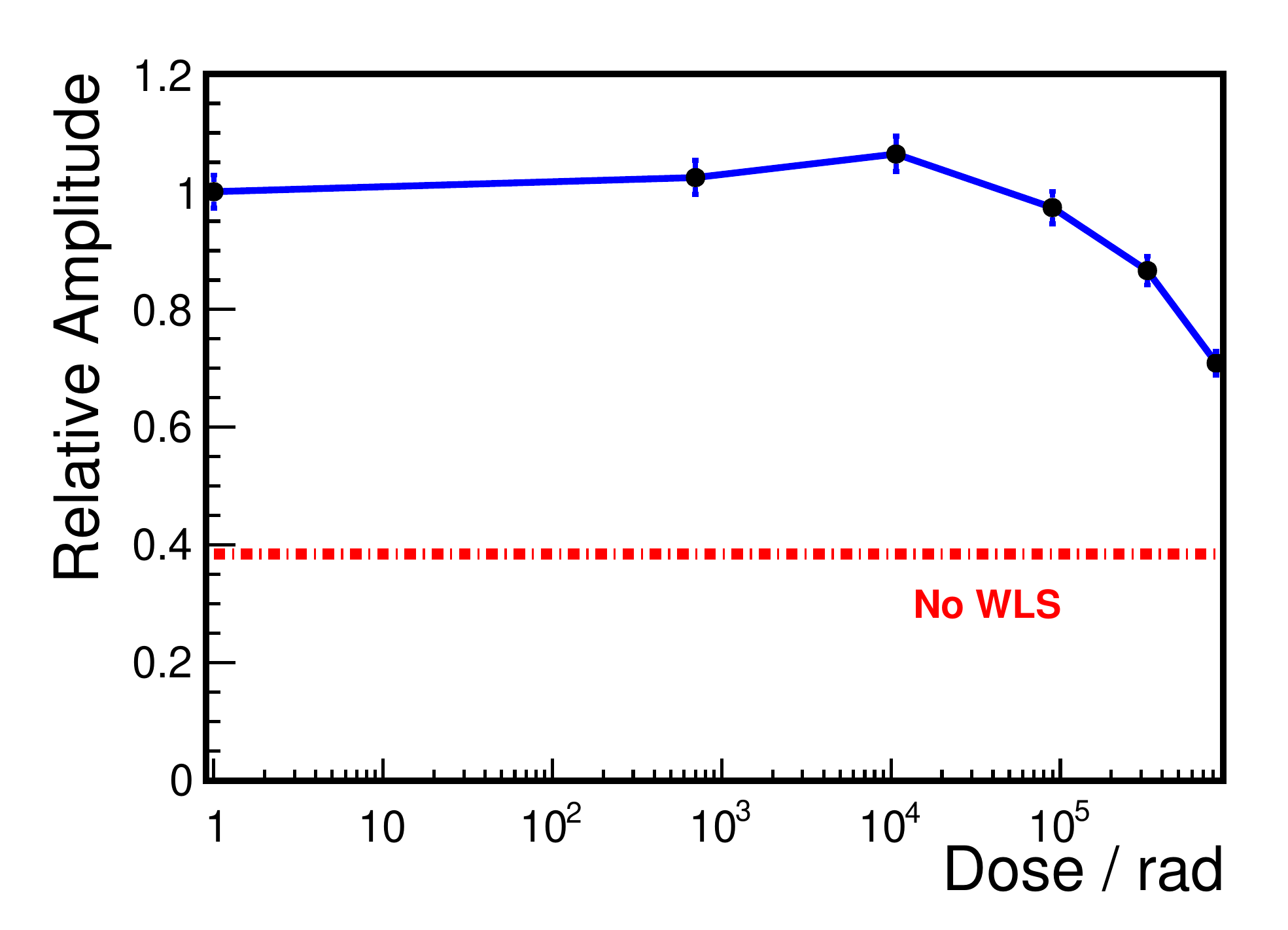}
	\caption{Relationship between the signal amplitude and TID dose. The dash-dotted red line shows the relative amplitude obtained from the unit without
		the NOL and in the absence of iradiation.}%
	\label{fig:dose_ly}
\end{figure}

In this test, the NOL composite film is irradiated up to a TID of around 1\:Mrad, where the radiation dose is calibrated by the
alanine/EPR standard dosimeter. After irradiation, the NOL composite film is assembled in the module as described in
Sec.~\ref{sec:cosmic_ray_test_of_the_wls_coating_emc_module} and the cosmic ray test is carried out to measure the change of the detected L.Y. In the
process of assembling the module, the condition of optical coupling between crystal and APD introduces $2\:\%$ uncertainty into
the signal amplitude.
Figure~\ref{fig:dose_ly} shows the relative signal amplitude of the cosmic ray under different TID.
The detected L.Y. does not change significantly with the TID up to 90\:krad and drops less than  $30\:\%$ under the TID of 833\:krad.
Therefore, the  NOL composite film is expected to  be resistant to radiation damage for the EMC of STCF~\cite{YANG2021165043}.

\begin{figure}[htpb]
	\centering
	\includegraphics[width=\linewidth]{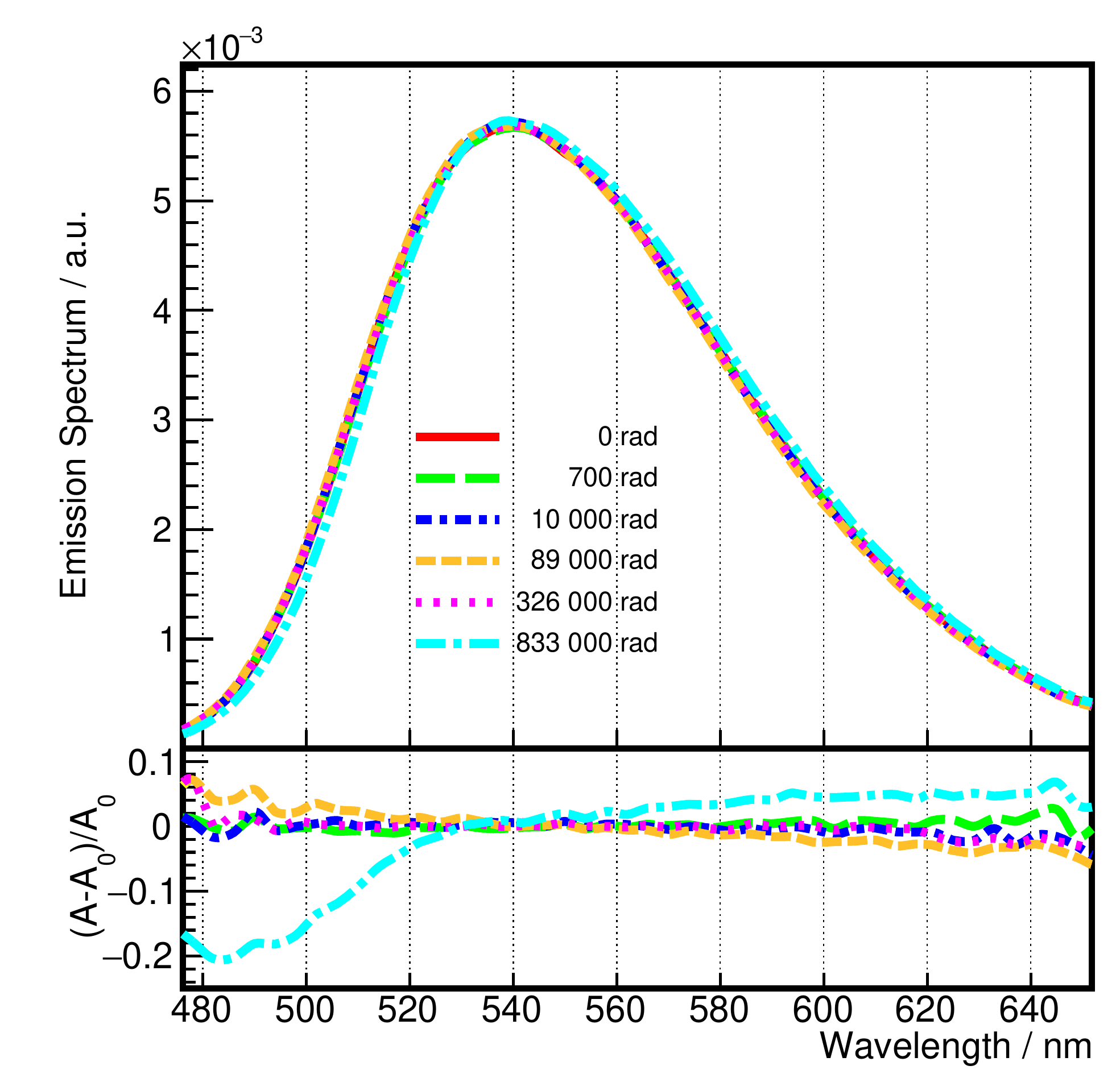}
	\caption{Emission spectrum of the NOL under varying TID dose. The spectrum is normalized by area and the lower plot shows the residual distribution of
		the spectrum. }%
	\label{fig:emission}
\end{figure}

To further study the properties of the NOL composite film, the emission spectrum of the NOL is reviewed under different TID by the HORIBA Fluorolog Tau-3 fluorometer.
The laser with wavelength 310\:nm is injected directly into the NOL composite film, and the emission spectrum is measured.
The emission spectrum of the NOL under different TID normalized by the area are shown in Fig.~\ref{fig:emission}, and the corresponding residual
distributions referring to that without irradiation are shown in the lower plot of  Fig.~\ref{fig:emission}.
No obvious change of the spectrum is observed up to a TID of 326\:krad, while the emission intensities at the small wavelength  region become smaller (up to
$20\:\%$ at wavelength 480\:nm) for that with the TID of 833\:krad.
Combining the results of signal amplitude measurement under different TIDs as shown in Fig.~\ref{fig:dose_ly},
the fluor-light production mechanism of the NOL is indicated to be stable and  unchanged with a TID up to  326\:krad, which is beyond the TID of
STCF EMC over the ten years of operation.
However, the emission spectrum variety with a TID above 833\:krad needs further study.

\section{Summary}%
\label{sec:summary}

In summary, the new generation electron-positron collider experiments in the high luminosity frontier require a crystal-based EMC with performances of fast
response, precision measurement, and high radiation
tolerance. This is challenging for manufacturing and operation. Various measurements are investigated to improve its performance, and increasing
the detected L.Y. is one of the most fundamental ones.
This paper explores the WLSP scheme with the novel NOL-coated composite film to improve the transmittance and Q.E. simultaneously for the undoped CsI-based EMC for STCF.
The detected L.Y. is proved to be effectively increased with a factor larger than 2 both by the MC simulation and the cosmic ray test.
Meanwhile, the radiation  tolerance of the composite NOL film is explored under $\gamma$-ray irradiation.
The results indicate the performance of the NOL film does not degrade with a TID up to 100\:krad, which fulfills the requirement of the
EMC of STCF for more than ten years of operation.
This paper's research provides an important approach and lays the foundation for the manufacture and operation of the EMC of STCF.

\section{Acknowledgments}%
\label{sec:acknowledgments}
The authors thank Hefei Comprehensive National Science Center for their support. This work was supported in part by National Key R\&D Program of China under
Contracts Nos. 2022YFA1602200 (2022YFA1602204); the international partnership program of
the Chinese Academy of Sciences under Grant No. 211134KYSB20200057 and by the Double First-Class university project foundation of USTC.


\begin{thebibliography}{10}
	\expandafter\ifx\csname url\endcsname\relax
		\def\url#1{\texttt{#1}}\fi
	\expandafter\ifx\csname urlprefix\endcsname\relax\def\urlprefix{URL }\fi
	\expandafter\ifx\csname href\endcsname\relax
		\def\href#1#2{#2} \def\path#1{#1}\fi

	\bibitem{Luo:2019gri}
	Q.~Luo, {Progress of Preliminary Work for the Accelerators of a 2-7GeV Super
			Tau Charm Facility at China}, in: {62nd ICFA Advanced Beam Dynamics Workshop
	on High Luminosity Circular $e^+ e^-$ Colliders}, 2019, p. TUOBB03.
	\newblock \href {https://doi.org/10.18429/JACoW-eeFACT2018-TUOBB03}
	{\path{doi:10.18429/JACoW-eeFACT2018-TUOBB03}}.

	\bibitem{snowmass}
	J.~B. Liu, X.~R. Lyu, R.~E. Mitchell, S.~L. Olsen, H.~P. Peng, Z.~G. Zhao,
	Y.~H. Zheng, X.~R. Zhou, {Physics Potential of a Super tau-Charm Facility},
	\url{https://www.snowmass21.org/docs/files/summaries/RF/SNOWMASS21-RF7_RF1_STCF-013.pdf}
	(2022).

	\bibitem{Prokhorova:2020jem}
	E.~Prokhorova, {Study of the fast calorimeter prototype for modern $e^+ e^-$
			factories}, JINST 15~(08) (2020) C08023.
	\newblock \href {https://doi.org/10.1088/1748-0221/15/08/C08023}
	{\path{doi:10.1088/1748-0221/15/08/C08023}}.

	\bibitem{Jin:2016mic}
	Y.~Jin, H.~Aihara, O.~V. Borshchev, D.~A. Epifanov, S.~A. Ponomarenko, N.~M.
	Surin, {Study of a pure CsI crystal readout by APD for Belle II end cap ECL
			upgrade}, Nucl. Instrum. Meth. A 824 (2016) 691--692.
	\newblock \href {https://doi.org/10.1016/j.nima.2015.07.034}
	{\path{doi:10.1016/j.nima.2015.07.034}}.

	\bibitem{luo2020design}
	L.~F. Luo, Z.~T. Shen, Y.~L. Huang, Z.~K. Jia, C.~Q. Feng, S.~B. Liu, {Design
			and optimization of the CSA-based readout electronics for STCF ECAL}, JINST
	15~(09) (2020) C09002.
	\newblock \href {https://doi.org/10.1088/1748-0221/15/09/C09002}
	{\path{doi:10.1088/1748-0221/15/09/C09002}}.

	\bibitem{Bedny:2009zz}
	I.~V. Bedny, A.~E. Bondar, V.~V. Cherepkov, D.~A. Epifanov, M.~G. Golkovsky,
	A.~S. Kuzmin, S.~B. Oreshkin, V.~E. Shebalin, B.~A. Shwartz, Y.~V. Usov,
	{Study of the radiation hardness of the pure CsI crystals}, Nucl. Instrum.
	Meth. A 598 (2009) 273--274.
	\newblock \href {https://doi.org/10.1016/j.nima.2008.08.106}
	{\path{doi:10.1016/j.nima.2008.08.106}}.

	\bibitem{8456600}
	F.~Yang, L.~Y. Zhang, C.~Hu, R.~Y. Zhu, {Slow Scintillation Component and
			Radiation-Induced Readout Noise in Undoped CsI Crystals}, IEEE Trans. Nucl.
	Sci. 65~(10) (2018) 2716--2723.
	\newblock \href {https://doi.org/10.1109/TNS.2018.2868678}
	{\path{doi:10.1109/TNS.2018.2868678}}.

	\bibitem{Kessler:1995xz}
	R.~S. Kessler, et~al., {Beam test of prototype CsI calorimeter}, Nucl. Instrum.
	Meth. A 368 (1996) 653--665.
	\newblock \href {https://doi.org/10.1016/0168-9002(95)00677-X}
	{\path{doi:10.1016/0168-9002(95)00677-X}}.

	\bibitem{Yamanaka:2012yma}
	T.~Yamanaka, {The J-PARC KOTO experiment}, PTEP 2012 (2012) 02B006.
	\newblock \href {https://doi.org/10.1093/ptep/pts057}
	{\path{doi:10.1093/ptep/pts057}}.

	\bibitem{Atanov:2017mvq}
	N.~Atanov, et~al., {Quality Assurance on Undoped CsI Crystals for the Mu2e
			Experiment}, IEEE Trans. Nucl. Sci. 65~(2) (2017) 752--757.
	\newblock \href {http://arxiv.org/abs/1802.08247} {\path{arXiv:1802.08247}},
	\href {https://doi.org/10.1109/TNS.2017.2786081}
	{\path{doi:10.1109/TNS.2017.2786081}}.

	\bibitem{10.1117/12.2187281}
	S.~A. Ponomarenko, N.~M. Surin, O.~V. Borshchev, M.~S. Skorotetcky, A.~M.
	Muzafarov, {Nanostructured organosilicon luminophores as a new concept of
			nanomaterials for highly efficient down-conversion of light}, in: S.~Cabrini,
	G.~L{\'e}rondel, A.~M. Schwartzberg, T.~Mokari (Eds.), Nanophotonic Materials
	XII, Vol. 9545, International Society for Optics and Photonics, SPIE, 2015,
	p. 954509.
	\newblock \href {https://doi.org/10.1117/12.2187281}
	{\path{doi:10.1117/12.2187281}}.

	\bibitem{skorotetcky2019synthesis}
	M.~Skorotetcky, O.~Borshchev, G.~Cherkaev, S.~Ponomarenko, Synthesis of
	nanostructured organosilicon luminophores based on phenyloxazoles, Russ. J.
	Org. Chem. 55~(1) (2019) 25--41.

	\bibitem{borshchev2017development}
	O.~Borshchev, A.~Cavalcante, L.~Gavardi, L.~Gruber, C.~Joram, S.~Ponomarenko,
	O.~Shinji, N.~Surin, Development of a new class of scintillating fibres with
	very short decay time and high light yield, JINST 12~(05) (2017) P05013.

	\bibitem{akimov2017test}
	D.~Y. Akimov, V.~Belov, O.~Borshchev, A.~Burenkov, Y.~L. Grishkin, A.~Karelin,
	A.~Kuchenkov, A.~Martemiyanov, S.~Ponomarenko, G.~Simakov, et~al., {Test of
			SensL SiPM coated with NOL-1 wavelength shifter in liquid xenon}, JINST
	12~(05) (2017) P05014.

	\bibitem{baxter2020coherent}
	D.~Baxter, J.~Collar, P.~Coloma, C.~Dahl, I.~Esteban, P.~Ferrario, J.~J.
	Gomez-Cadenas, M.~Gonzalez-Garcia, A.~Kavner, C.~Lewis, et~al., Coherent
	elastic neutrino-nucleus scattering at the european spallation source, JHEP
	2020~(2) (2020) 1--38.

	\bibitem{luminotech}
	Lum{I}nno{T}ech {LLC}, \url{https://www.luminnotech.com/}.

	\bibitem{ren}
	G.~H. Ren, Z.~H. Song, Z.~C. Zhang, K.~Zhang, F.~Yang, H.~Y. Li, F.~Chen,
	{Luminescence and Decay Time Properties of Pure CsI Crystals}, J. Inorg.
	Mater 32 (2017) 169.
	\newblock \href {https://doi.org/10.15541/jim20160304}
	{\path{doi:10.15541/jim20160304}}.

	\bibitem{hamamatsu}
	{HAMAMATSU} {S}866-1010 short wavelength type {APD},
	\url{https://www.hamamatsu.com/us/en/product/optical-sensors/apd/si-apd/S8664-1010.html}.

	\bibitem{GEANT4:2002zbu}
	S.~Agostinelli, et~al., {GEANT4--a simulation toolkit}, Nucl. Instrum. Meth. A
	506 (2003) 250--303.
	\newblock \href {https://doi.org/10.1016/S0168-9002(03)01368-8}
	{\path{doi:10.1016/S0168-9002(03)01368-8}}.

	\bibitem{Allison:2006ve}
	J.~Allison, et~al., {Geant4 developments and applications}, IEEE Trans. Nucl.
	Sci. 53 (2006) 270.
	\newblock \href {https://doi.org/10.1109/TNS.2006.869826}
	{\path{doi:10.1109/TNS.2006.869826}}.

	\bibitem{Allison:2016lfl}
	J.~Allison, et~al., {Recent developments in Geant4}, Nucl. Instrum. Meth. A 835
	(2016) 186--225.
	\newblock \href {https://doi.org/10.1016/j.nima.2016.06.125}
	{\path{doi:10.1016/j.nima.2016.06.125}}.

	\bibitem{doi:10.1063/1.555536}
	H.~H. Li, Refractive index of alkali halides and its wavelength and temperature
	derivatives, J. Phys. Chem. Ref. Data 5~(2) (1976) 329--528.
	\newblock \href {https://doi.org/10.1063/1.555536}
	{\path{doi:10.1063/1.555536}}.

	\bibitem{Ikagawa:2003md}
	T.~Ikagawa, et~al., {Performance of large area avalanche photodiode for
			low-energy X-rays and gamma rays scintillation detection}, Nucl. Instrum.
	Meth. A 515 (2003) 671--679.
	\newblock \href {https://doi.org/10.1016/j.nima.2003.07.024}
	{\path{doi:10.1016/j.nima.2003.07.024}}.

	\bibitem{6168236}
	M.~Janecek, Reflectivity spectra for commonly used reflectors, IEEE Trans.
	Nucl. Sci. 59~(3) (2012) 490--497.
	\newblock \href {https://doi.org/10.1109/TNS.2012.2183385}
	{\path{doi:10.1109/TNS.2012.2183385}}.

	\bibitem{Roncali_2017}
	E.~Roncali, M.~Stockhoff, S.~R. Cherry, An integrated model of
	scintillator-reflector properties for advanced simulations of optical
	transport, Phys. Med. Biol. 62~(12) (2017) 4811--4830.
	\newblock \href {https://doi.org/10.1088/1361-6560/aa6ca5}
	{\path{doi:10.1088/1361-6560/aa6ca5}}.

	\bibitem{4437209}
	C.~Hagmann, D.~Lange, D.~Wright, Cosmic-ray shower generator (cry) for monte
	carlo transport codes, in: 2007 IEEE Nuclear Science Symposium Conference
	Record, Vol.~2, 2007, pp. 1143--1146.
	\newblock \href {https://doi.org/10.1109/NSSMIC.2007.4437209}
	{\path{doi:10.1109/NSSMIC.2007.4437209}}.

	\bibitem{Luo:2022rjf}
	L.~F. Luo, Z.~K. Jia, Z.~T. Shen, Y.~L. Zhang, S.~B. Liu, {Study on time
			measurement for CSA-based readout electronics in STCF ECAL}, JINST 17~(02)
	(2022) P02034.
	\newblock \href {https://doi.org/10.1088/1748-0221/17/02/P02034}
	{\path{doi:10.1088/1748-0221/17/02/P02034}}.

	\bibitem{Aihara:2016ct}
	H.~Aihara, O.~Borshchev, D.~Epifanov, Y.~Jin, S.~A. Ponomarenko, N.~M. Surin,
	{Study of scintillation counter consisting of a pure {C}s{I} crystal and
		{APD}}, PoS PhotoDet2015 (2016) 052.
	\newblock \href {https://doi.org/10.22323/1.252.0052}
	{\path{doi:10.22323/1.252.0052}}.

	\bibitem{jiao2021effect}
	L.~M. Jiao, Y.~Wang, Z.~H. Wu, H.~Shen, H.~Q. Weng, H.~B. Chen, W.~Huang, M.~Z.
	Wang, X.~W. Ge, M.~Z. Lin, Effect of gamma and neutron irradiation on
	properties of boron nitride/epoxy resin composites, Polym. Degrad. Stab. 190
	(2021) 109643.

	\bibitem{YANG2021165043}
	Y.~Fan, C.~Hu, L.~Y. Zhang, R.~Y. Zhu, {UV}–{V}isible reflectance of common
	light reflectors and their degradation after an ionization dose up to 100
		{M}rad, Nucl. Instrum. Meth. A 992 (2021) 165043.
	\newblock \href {https://doi.org/https://doi.org/10.1016/j.nima.2021.165043}
	{\path{doi:https://doi.org/10.1016/j.nima.2021.165043}}.

\end{thebibliography}

\end{document}